\documentclass[leqno]{article}%
\usepackage{amsfonts,amsmath,amsthm,amssymb,times}
\usepackage{graphicx}
\usepackage{amsmath}
\usepackage{amsfonts}
\usepackage{amssymb}%
\newtheorem{theorem}{THEOREM}[section]
\newtheorem{lemma}{LEMMA} [section]
\newtheorem{definition}{DEFINITION}[section]
\newtheorem{corollary}{COROLLARY}[section]
\newtheorem{problem}[theorem]{PROBLEM}
\newtheorem{remark}{REMARK}[section]

\theoremstyle{definition}

\begin{document}

\title{A NONPARAMETRIC APPROACH TO
3D SHAPE ANALYSIS FROM DIGITAL CAMERA IMAGES - I.\\ In Memory of
W.P. Dayawansa.}
\author{V. Patrangenaru $^{1}$\thanks{Research supported by
National Science Foundation Grant DMS-0652353 and by National
Security Agency Research Grant H98230-08-1-0058}, X. Liu $^{1}$
\thanks{Research supported by National Science Foundation Grants
CCF-0514743 and DMS-0713012},
 S. Sugathadasa $^{2}$ \\ $^{1}$ {\small Florida State
University}\\ $^{2}$ {\small Texas Tech University}}
\date{June, 5, 2008} \maketitle
\begin{abstract}
In this article, for the first time, one develops a nonparametric
methodology for an analysis of shapes of configurations of
landmarks on real 3D objects from regular camera photographs, thus
making 3D shape analysis very accessible. A fundamental result in
computer vision by Faugeras (1992), Hartley, Gupta and Chang
(1992) is that generically, a finite 3D configuration of points
can be retrieved up to a projective transformation, from
corresponding configurations in a pair of camera images.
Consequently, the projective shape of a 3D configuration can be
retrieved from two of its planar views. Given the inherent
registration errors, the 3D projective shape can be estimated from
a sample of photos of the scene containing that configuration.
Projective shapes are here regarded as points on projective shape
manifolds. Using large sample and nonparametric bootstrap
methodology for extrinsic means on manifolds, one gives confidence
regions and tests for the mean projective shape of a 3D
configuration from its 2D camera images.
\end{abstract}

\numberwithin{equation}{section}
\section{Introduction}
Until now, statistical analysis of similarity shapes from images
was restricted to a small amount of data, since similarity shape
appearance is relative to the camera position with respect to the
scene pictured.\\
In this paper, for the first time, we study the shape of a 3D
configuration from its 2D images in photographs of the
configuration, without any camera positioning restriction relative
to the scene pictured. Our nonparametric methodology is manifold
based, and uses standard reconstruction methods in computer vision. \\
In absence of occlusions, a set of point correspondences in two
views can be used to retrieve the 3D configuration of points.
Faugeras (1992) and Hartley et. al. (1992) state that two such
reconstructions differ by a projective transformation in 3D.
Sughatadasa (2006) and Patrangenaru and Sughatadasa (2006) noticed
that actually the object which is recovered without ambiguity is
the projective shape of the configuration, which casts a new light
on the role of projective
shape in the identification of a spatial configuration.\\
Projective shape is the natural approach to shape analysis from
digital images, since the vast majority of libraries of images are
acquired via a central projection from the scene pictured to the
black box recording plane. Hartley and Zisserman (2004, p.1) note
that "this often rends classical shape analysis of a spatial scene
impossible, since similarity is not preserved when a camera is
moving." \\
Advances in statistical analysis of projective shape have been
slowed down due to overemphasis on the importance of similarity
shape in image analysis, with little focus on the principles of
image acquisition or binocular vision. Progress was also affected
by lack of a geometric model for the space of projective shapes,
and ultimately probably by insufficient dialogue between
researchers
in geometry,computer vision and statistical shape analysis.\\
For reasons presented above, projective shapes have been studied
only recently, and except for one concrete 3D example due to
Sughatadasa(2006), to be found in Liu et al. (2007), the
literature was bound to linear or planar projective shape
analyzes. Examples of 2D projective shape analysis can be found in
Maybank (1994), Mardia et. al. (1996), Goodall and Mardia (1999),
Patrangenaru (2001), Lee et. al. (2004), Paige et. al. (2005),
Mardia and Patrangenaru (2005), Kent and Mardia (2006, 2007) and
Munk et. al.
(2007). \\
Our main goal here is to derive a natural concept of 3D shape that
can be extracted from data recorded from camera images. The
statistical methodology for estimation of a mean 3D projective
shape is nonparametric, based on large sample theory and on
Efron's bootstrap ( Efron (1979, 1982)). In this paper, a 3D
projective shape is regarded as a random object on a projective
shape space. Since typically samples of images are small, in order
to estimate the mean projective shape we use nonparametric
bootstrap for the studentized sample mean projective shape on a
manifold, as shown in Bhattacharya and Patrangenaru (2005). This
bootstrap distribution was essentially presented in Mardia and
Patrangenaru (2005). Since, while running the projective shape
estimation algorithm in Mardia and Patrangenaru (2005) on a
concrete data set, Liu et al. (2007) have found typos in some
formulas. In this paper we are making the necessary corrections. \\
In section 2 we present projective geometry concepts and facts
that are needed in section 3, such as projective space, projective
frames, and projective coordinates. We also introduce computer
vision concepts, such as essential matrix and fundamental matrix ,
associated with a pair of camera views of a 3D scene that is
needed in the reconstruction of that scene from 2D calibrated, and
respectively non-calibrated camera images. We then state the
Faugeras-Hartley-Gupta-Chang projective ambiguity theorem for the
scene reconstructed from two non-calibrated camera views.
 For the reconstruction of a configuration of points in space from its
views in a pair of images, we refer to a computational algorithm in
Ma
et. al. (2006). \\
In Section 3 we introduce projective shapes of configurations of
points in $\mathbb{R}^m$ or in $\mathbb{R}P^m,$ and the
multivariate axial geometric model for the projective shape space,
which is our choice for a statistical study of projective shape.
The Faugeras-Hartley-Gupta-Chang theorem is reformulated in
Theorem \ref{key_result} in terms of projective shapes: {\it if
$\mathcal{R}$ is a 3D reconstruction of a spatial configuration
$\mathcal{C}$ from two of its uncalibrated camera views, then
$\mathcal{R}$ and $\mathcal{C}$ have the same projective shape}.
This is the key result for our projective shape analysis of
spatial configurations, which opens the Statistical Shape Analysis
door to Computer Vision and Pattern
Recognition of 3D scenes. \\
Since projective shape spaces are identified via projective frames
with products of axial spaces, in section 4 we approach
multivariate axial distributions via a quadratic equivariant
embedding of a product of $q$ copies of $\mathbb{R}P^m$ in
products of spaces of symmetric matrices. A theorem on the
asymptotic distributions of extrinsic sample means of multivariate
axes, stated without proof and with some minor typos in Mardia and
Patrangenaru (2005) is given in this section (Theorem
\ref{asymptprsh}) with a full proof. The corrections to Mardia and
Patrangenaru (2005) are listed in Remark \ref{corrections}. The
asymptotic and nonparametric bootstrap distribution results are
used to derive confidence regions for extrinsic mean projective
shapes. If a random projective shape has a nondegenerated
extrinsic covariance matrix, one may studentize the extrinsic
sample mean to generate asymptotically chi square distributions
that are useful for large sample confidence regions in Corollary
\ref{largeconfprsh}, or nonparametric bootstrap confidence regions
if the sample is small in Corollary \ref{onsamplesmall}. If the
extrinsic covariance matrix is degenerated, and the axial
marginals have nondegenerated extrinsic covariance matrices, one
gives a Bonferroni type of argument for axial marginals to derive
confidence regions for the mean projective shape in Corollary
\ref{onsamplesmall-s}.
\section{Basic Projective Geometry for Ideal Pinhole Camera
Image Acquisition} Pinhole camera image acquisition is based on a
central projection from the 3D world to the 2D photograph.
Distances between observed points are not proportional to
distances between their corresponding points in the photograph,
and Euclidean geometry is inappropriate to model the relationship
between a 3D object and its picture, even if the object is flat.
The natural approach to ideal pinhole camera image acquisition is
via projective geometry, which also provides a logical
justification for the mental reconstruction of a spatial scene
from binocular retinal images, playing a central role in human
vision. In this section we review some of the basics of projective
geometry that are useful in understanding  image formation and
scene retrieval from ideal pinhole camera images.
\subsection{Basics of Projective Geometry}
Consider a real vector space $V.$ Two vectors $x, y \in V
\backslash \{0_V \}$ are equivalent if they differ by a scalar
multiple. The equivalence class of $ x \in V \backslash \{0_V  \}$
is labeled $[ x ],$ and the set of all such equivalence classes is
the {\it projective space} $P(V)$ associated with $V,$ $P(V) = \{
[x], x \in V \backslash O_V \}.$ The real projective space
$\mathbb R P^m$ is $P(\mathbb R^{m+1}).$ Another notation for a
{\it projective point} $ p = [ x ] \in \mathbb R P^m, $
equivalence class of $ x = (x^1, \dots, x^{m+1}) \in \mathbb
R^{m+1}, p = [ x^1: x^2: \dots : x^{m+1} ],$ features the
 {\it homogeneous coordinates} $(x^1, \dots, x^{m+1})$ of $p,$
 which are determined up to a multiplicative constant. A projective
 point $ p $ admits also a {\it spherical representation }, when thought
 of as a pair of antipodal points on the $m$ dimensional unit sphere, $p = \{ z , - z \},
x = (x^1, x^2, \dots, x^{m+1}), (x^1)^2 + \dots +(x^{m+1})^2=1.$
A $d$ - dimensional {\it projective subspace} of $\mathbb{R}P^m$ is
a projective space $P(V)$, where $V$ is a $(d+1)$-dimensional vector
subspace of $\mathbb R^{m+1}$. A codimension one projective subspace
of $\mathbb R P^m$ is also called {\it hyperplane.} The {\it linear
span} of a subset $D$ of $\mathbb RP^m $ is the smallest projective
subspace of $\mathbb{R} P^m$ containing $D.$ We say that $k$ points
in $\mathbb R P^m$ are in {\it general position} if their linear
span is $\mathbb R P^m.$ If $k$
points in $\mathbb R P^m$ are in general position, then $k \geq m+1.$ \\
The numerical space $\mathbb R^m$ can be {\it embedded} in $\mathbb
R P^m$, preserving collinearity. An example of such an {\it affine}
embedding is
\begin{eqnarray} \label{homo} h((u^1,...,u^m)) = [u^1:~~...~~:u^m:1] =
[\tilde{u}],
\end{eqnarray}
where $\tilde{u} = (u^1,\dots, u^m, 1)^T,$ and in general, an affine
embedding is given for any $A \in Gl(m+1, \mathbb R),$ by $ \ h_A(u)
= [A\tilde{u}].$
The complement of the range of the embedding $h$ in \eqref{homo}
is the hyperplane $\mathbb{R}P^{m-1},$ set of points
$[x^1:\dots: x^m:0] \in \mathbb RP^m.$\\
Conversely, the {\it inhomogeneous ( affine ) coordinates } $(
u^1,\dots ,u^m) $ of a point $ p = [ x^1: x^2: \dots : x^{m+1} ] \in
\mathbb R P^m \backslash \mathbb{R}P^{m-1}$ are given by
\begin{equation} \label{inhomogeneous} u^j = \frac{x^j}{x^{m+1}}, \forall j = 1, \dots, m.
\end{equation}
Consider now the linear transformation from $\mathbb R^{m'+1}$ to
$\mathbb R^{m +1}$ defined by the matrix $B \in M(m+1,m'+1;\mathbb
R)$ and its kernel $ K = \{ x \in \mathbb R^{m'+1}, Bx = 0\}.$ The
{\it projective map $\beta:\mathbb RP^{m'} \backslash P(K) \to
\mathbb RP ^m,$ associated with $B$} is defined by
\begin{equation} \label{beta}
\beta ([x]) = [Bx].\end{equation} In particular, a {\it projective
transformation} $\beta$ of $\mathbb{R} P^m$ is the projective map
associated with a
 nonsingular matrix $B \ \in GL (m + 1, \mathbb R)$ and its action
on $\mathbb{R}P^m:$
\begin{eqnarray} \label{alpha} \beta ([x^1: \dots :x^{m + 1}])
= [B(x^1, \dots , x^{m + 1})] .
\end{eqnarray}
In affine coordinates ( inverse of the affine embedding
\eqref{homo}), the projective transformation \eqref{alpha} is given
by $v = f(u)$, with
\begin{equation} \label{proje}
v^j = \frac{a_{m+1}^j + \sum_{i=1}^m a_i^ju^i}{ a_{m+1}^{m+1} +
\sum_{i=1}^m a_i^{m+1} u^i}, \forall j=1, \dots, m
\end{equation}
where $\det B = \det((a_i^j)_{i,j=1,\dots,m+1}) \neq 0.$
An {\it affine} transformation of $\mathbb RP^m, v = A u + b, A \in
GL(m,\mathbb R), t \in \mathbb R^m, $ is a particular case of
projective transformation $\alpha,$ associated with the matrix $B \
\in GL (m + 1, \mathbb R), $ given by
\begin{equation} \label{affine-transf}
B = \left (\begin{matrix} A & b \cr 0_m^T & 1 \cr
\end{matrix}\right).
\end{equation}
A {\it projective frame} in an $m$ dimensional projective space (
or {\it projective basis } in computer vision literature, see e.g.
Hartley (1993)) is an ordered set of $m + 2$ projective points in
general position. An example of projective frame in
$\mathbb{R}P^m$ is the {\it standard projective frame}
$([e_1],\dots,[e_{m +
1}],[e_1 + ... + e_{m + 1}]).$ \\
In projective shape analysis it is preferable to employ coordinates
invariant with respect to the group PGL$(m)$ of projective
transformations. A projective transformation takes a projective
frame to a projective frame, and its action on $\mathbb R P^m$ is
determined by its action on a projective frame, therefore if we
define the {\it projective coordinate(s)} of a point $p \in
\mathbb{R}P^m$ w.r.t. a projective frame  $\pi = (p_1,\dots, p_{m +
2})$ as being given by
\begin{equation} \label{projcoord}
p^{ \pi} = \beta^{-1} (p), \end{equation} where $ \beta \in
PGL(m)$ is a projective transformation taking the standard
projective frame to $\pi.$ These coordinates have automatically
the invariance property.
\begin{remark} Assume $u, u_1, \dots, u_{m+2}$
are points in $\mathbb{R}^{m},$ such that $\pi = ( [\tilde{u}_1],
\dots, [\tilde{u}_{m+2}])$ is a projective frame. If we consider the
$(m+1) \times (m+1) $ matrix $U_m = [\tilde{u}_1^T, \dots,
\tilde{u}_{m+1}^T],$ the projective coordinates of $ p =[\tilde{u}]$
w.r.t. $\pi$ are given by
\begin{equation} \label{projcoord_formula}
 p^{\pi} = [ y^1(u): \dots: y^{m+1}(u)],
\end{equation}
where
\begin{eqnarray} \label{v}
v(u) = U_m^{-1} \tilde{u}^T
\end{eqnarray}
and
\begin{equation} \label{y}
y^j(u)=\frac{v^j(u)}{v^j(u_{m+2})}, \forall j = 1, \dots, m+1.
\end{equation}
\end{remark}
Note that in our notation, the superscripts are reserved for the
components of a point, whereas the subscripts are for the labels
of points.  The projective coordinate(s) of $x$ are given by the
point $[z^1(x):\dots :z^{m+1}(x)] \in \mathbb{R}P^m.$

\subsection{Projective geometry and image acquisition in ideal digital
cameras.} An introduction to the geometry pinhole camera principle
can be found in 3D-Vision texts including Ma et. al. (2006),
Hartley and Zisserman (2004) \cite{ziss}, Birchfeld (1998)
\cite{Birch}, etc. In this section we give such a description in
our projective geometry notation.  {\it Ideal pinhole camera}
image acquisition can be thought of in terms of a central
projection $\beta: \mathbb RP^3 \backslash \mathbb RP^2 \to
\mathbb RP^2,$ whose representation in conveniently selected
affine coordinates $(x, y, z) \in \mathbb R^3, \ (u,v) \in
\mathbb{R}^2$ is given by
\begin{eqnarray} \label{ideal_camera}
u = f\frac{x}{z} \nonumber \\
v = f\frac{y}{z},
\end{eqnarray}
where $f$ is the {\it focal length,} i.e. the distance from the
{\it image sensor} or film to the pinhole or {\it principal plane
of the lens}
 $\mathbb RP^2,$ which is the
complement of the domain of $\beta$ in $\mathbb RP^3.$ In
homogeneous coordinates $[x:y:z:w], [u:v:t]$ the {\it perspective
projective map} $\beta$ can be represented by the matrix $B \in
M(3,4;\mathbb R)$ given by:
\begin{equation} \label{ideal-cam-proj}B = \left (\begin{matrix} f & 0 & 0 & 0
\cr 0 & f & 0 & 0 \cr 0 & 0 & 1 & 0 \cr
\end{matrix}\right). \end{equation}
Digital cameras image acquisition is based on a slightly different
projective transformation, that in addition takes into account
internal camera parameters such as {\it pixel aspect ratio,
skewness parameter and principal point (origin of image
coordinates in the principal plane).} For such cameras, the
projective map \eqref{ideal-cam-proj} is altered by a composition
with a matrix accounting for camera internal calibration
parameters. If we also take into consideration the change of
coordinates between the initial and current camera position
involving a roto-translation $(R, t) \in SO(3)\times
\mathbb{R}^3,$ the projective map of a pinhole camera image
acquisition $\tilde \pi$ is associated with the matrix:
\begin{equation} \label{camera_par}
\tilde{B} = C_{\text{int}} B G = \left(\begin{matrix} k_u & k_c &
u_0 \cr 0 & k_v & v_0 \cr 0 & 0 & 1
\end{matrix}\right) \left(\begin{matrix} f & 0 & 0 & 0 \cr 0 & f & 0 & 0 \cr 0 & 0 &
1& 0
\end{matrix} \right) \left (\begin{matrix} R & t \cr 0_3^T & 1 \cr
\end{matrix}\right) =  A G,
  \end{equation}
where $k_u$ and $k_v$ are scale factors of the image plane in units
of the focal length $f,$ and $\theta = cot^{-1}k_c$ is the skew, and
$(u_0,v_0)$ is the principal point. The matrix $A$ contains the
internal parameters and the projection map \eqref{ideal-cam-proj},
while $E$ contains the external parameters. The columns of the
matrix $\tilde B $ are the columns of a $3 \times 3$ matrix $P$
followed by a $3 \times 1$ vector p:
\begin{equation}
\tilde B = \left( \begin{smallmatrix}P & p \end{smallmatrix} \right)
\end{equation}
so that
\begin{equation} \label{decomposition}
P = AR \; \; \mbox{and} \; \; p = A t.
\end{equation}
\subsection{Essential and fundamental matrices}
Consider now two positions of a camera directed at a point $[u]
\in \mathbb RP^3,$ and the projective points associated with its
images taken at these locations of the camera, $m_a=[u_a]\in
\mathbb RP^2, a =1,2,$ where $u_1, u_2 \in \mathbb R^3 \backslash
\{0\}.$ If we assume the camera's internal parameters are known (
camera is {\it calibrated),} then, with respect to the camera's
coordinates
frame at each position, we may assume $C_{\text{int}} =I_3.$\\
Since the lines joining the two locations of the camera optical
center with the image points meet at  $[u],$ these two lines and
the line joining the two locations of the camera optical center
are coplanar. The plane containing these lines is the
{\it epipolar plane} associated with the point $[u].$\\
Assume we refer all the points to one coordinate system, say the
coordinate system of the second position of the camera. The
position vectors of first and second image points are $t + Ru_1, $
respectively  $ u_2,$ and and the vector from one optical center
to the other is $t.$ Here the change of  coordinates between the
Euclidean frames corresponding to the two camera positions is
given by a roto-translation $(R,t) \in SO(3)\times \mathbb R^3.$
The three vectors above are directions of lines in the epipolar
plane, therefore
\begin{equation} \label{essential-pre1}u_2^T ( t \times ( R  u_1)) =
0.
\end{equation}
By defining $t_{\times}$ as the matrix associated with the linear
operator $ y \to t\times  y $ we can rewrite the equation
\eqref{essential-pre1} as follows
\begin{equation}\label{essential}
{u}_2^T (t_{\times} R {u}_1)) = {u}_2^T E {u}_1 = 0,
\end{equation}
where $E = t_{\times} R$ is the so called {\it essential matrix}.\\
If the camera is uncalibrated, then the matrices $A_1 = A_2 = A$ in
\eqref{decomposition} containing the camera internal parameters,
yield the homogeneous pixel coordinates:
\begin{eqnarray} \label{ves} {v}_1 & = & A {u}_1 \\
{v}_2 & = & A {u}_2.
\end{eqnarray}
Thus:
\begin{equation} \label{fundamental-pre}
(A^{-1} { v}_2)^T({t}\times R A^{-1} {v}_1) = {v}_2^T
A^{-1}({t}\times RA^{-1}{v}_1) = 0,
\end{equation}
and we obtain
\begin{equation}\label{fundamental}
{v}_2^T F {v}_1= 0,
\end{equation}
where $F = (A^{-1})^T E A^{-1},$ with $E$ the essential matrix in
\eqref{essential} is the so called {\it fundamental matrix}. The
fundamental matrix depends only on the relative position of the
two cameras, and on their internal parameters. It has rank two,
depending on seven real constants.

\subsection{Reconstruction of a 3D scene from two of its 2D images.}
If we select conveniently the coordinates for the first camera
position, and also incorporating  the internal parameters, we may
assume that the matrix associated with $\tilde \beta_1$ in
equations \eqref{beta} and \eqref{camera_par} is $B_1 = (I | 0),$
and the fundamental matrix factors as follows : $F = t_{\times}
R,$ where $ B_2 = (R | t)$ is the matrix defining $\tilde \beta_2$
( see equations \eqref{beta} and \eqref{camera_par}). Note that
here
 $R$ is nonsingular, and it does not
necessarily represent the matrix of a rotation. Let $[v_1], [v_2]
\in \mathbb RP^2$ be given by \eqref{ves} associated with a pair
$[u_1], [u_2] \in \mathbb RP^2$ corresponding to matched points in
two images. We seek a point $[u] \in \mathbb RP^3$ such that
$[v_i] = \tilde \beta_i [u], i =1,2. $ From the relation $v_2^T F
v_1 = v_2^T t_{\times}R v_1 = v_2^T(t \times R v_1) = 0,$ it
follows that $v_2, R v_1, t$ are linearly dependent and we may
assume that $R v_1 = b v_2 - a t.$ Moreover, since $v_1$ is
defined up to a scalar multiple, we may assume that $R v_1 = v_2 -
a t,$ and define $[u] \in \mathbb RP^3$ by $u = (v_1^T, a)^T.$ Now
$B_1 u = (I | 0)u = v_1,$ and $B_2 u = (R | t)u = Rv_1 + at =
v_2,$ therefore $[u]$ is a desired solution to the reconstruction
problem. As shown, $[u]$ is determined by the two camera
projection matrices $B_1$ and $B_2.$ If we choose a different pair
of camera matrices $B_1 H$ and $B_2 H$ yielding the same
fundamental matrix $F,$ then, in order to preserve the same pair
of matched image points, the point $[u]$ must be replaced by
$[H^{-1}u].$
\begin{problem} \label{noncalibrated}
The problem of the reconstruction of a configuration of points in
3D from two ideal uncalibrated camera images, is equivalent to the
following: given two camera images $\mathbb RP^2_1, \mathbb
RP^2_2$ of unknown relative position and unknown internal camera
parameters, and two matching sets of labelled points $\{p_{a,1},
\dots, p_{a,k}\}\subset \mathbb RP^2_a, a =1,2,$ find a
configuration of points $p_1, \dots, p_k \in \mathbb R P^3$ such
that there exist two positions of the camera $\mathbb RP^2_1,
\mathbb RP^2_2$ for which $\tilde \beta_a (p_j) = p_{a,j}, \forall
a = 1,2, j = 1, \dots, k.$
\end{problem}
The above discussion proves the following theorem  (Faugeras(1992),
Hartley et al.(1992)):
\begin{theorem}\label{reconstruction}
The reconstruction problem for two non calibrated camera images
has a solution in terms of the fundamental matrix $F = t_{\times}
R. $ Any two solutions can be obtained from each other by a
projective transformation in $\mathbb{R}P^3.$
\end{theorem}
\begin{remark}
Note that, although the configurations in correspondence are
finite, their size is arbitrarily large, and the assumption of
finite matching labelled pairs can be replaced by an assumption of
parameterized sets in correspondence. Therefore, in absence of
occlusions, a 3D configuration can be reconstructed from 2D
images, and this reconstruction is unique up to a projective
transformation.
\end{remark}
\subsection{ Estimation of the fundamental matrix.} Since equation
\eqref{fundamental} is homogeneous as a linear equation in $F,$
and $F$ has rank two, this matrix depends on seven independent
parameters. Therefore, in principle, $F$ can be recovered from
corresponding configurations of seven points. Due to the fact that
the nature of digital imaging data is inherently discrete and
errors occur also in landmark registration, $F$ can be estimated
using configurations of eight or more points $p_{a,i}, a=1,2, i =
1, \dots k, k \geq 8,$ whose stacked homogeneous coordinates are
the $k \times 3 $ matrices $y_a, a =1, 2.$ The linear system for
$F$ is
\begin{equation}\label{fundamental_system-estimate}
    y_2^T F y_1 = 0
\end{equation}
and can be written as
\begin{equation}
    f^T Y= 0,
\end{equation}
where $f$ is a vectorized form of $F.$ \\
A refined eight point algorithm for the estimate $\hat{F}$ of the
fundamental matrix $F$ can be found in Ma et al. (2006, p. 188, p.
395).
\section{ Projective Shape and 3D Reconstruction }
\begin{definition} \label{prsh}
 Two configurations of points in $\mathbb{R}^m$ have the
same {\it projective shape} if they differ by a projective
transformation of $\mathbb{R}^m.$
\end{definition}%
 Unlike similarities or affine transformations, projective
transformations of $\mathbb{R}^m$ do not have a group structure
under composition of maps( the domain of definition of the
composition of two such maps is smaller than the maximal domain of
a projective transformation in $\mathbb{R}^m ).$ To avoid this
complication, rather than considering the projective shapes of
configurations in $\mathbb{R}^m,$ we consider projective shapes of
configurations in $\mathbb{R}P^m.$ A {\it projective shape} of a
k-ad ( configuration of k landmarks or labelled points ) is the
orbit of that k-ad under projective transformations with respect
to the diagonal action
\begin{equation} \label{action}
\alpha_k(p_1, \dots, p_k) = (\alpha(p_1), \dots, \alpha, (p_k)).
\end{equation}
Since the action \eqref{alpha} of $\beta \in PGL(m)$ on $ [x] \in
\mathbb{R}P^m,$ when expressed in inhomogeneous coordinates
\eqref{inhomogeneous}, reduces to \eqref{proje}, if two
configurations $\Gamma_1, \Gamma_2$ of points in $\mathbb{R}^m$
have the same projective shape, then $h(\Gamma_1), h(\Gamma_2)$
have the same projective shape in $\mathbb{R}P^m$ ( $h$ is the
affine embedding
given by \eqref{homo}). \\
Patrangenaru (1999, 2001) considered the set $ G(k,m)$ of k-ads
$(p_1,...,p_k), k > m + 2,$ for which $\pi = (p_1,...,p_{m+2})$ is
a projective frame. $PGL(m)$ acts simply transitively on $G(k,m)$
and the projective shape space $P\Sigma^k_m$, is the quotient
$G(k,m)/PGL(m)$. Using the projective coordinates $(p_{m+3}^{
\pi}, \dots, p_{k}^{ \pi})$ given by  \eqref{projcoord} one can
show that $P\Sigma^k_m$ is a manifold diffeomorphic with $(\mathbb
R P^m)^{k-m-2}.$ The projective frame representation has two
useful features: firstly, the projective shape space has a
manifold structure, thus allowing the use of the asymptotic theory
for means on manifolds in Bhattacharya and Patrangenaru (2003,
2005). Secondly, it can be extended to infinite dimensional
projective shape spaces, such as projective shapes of curves, as
shown in Munk et al. (2007). This approach has the advantage of
being inductive in the sense that each new landmark of a
configuration adds an extra marginal axial coordinate, thus
allowing to detect its overall contribution to the variability of
the configuration, as well as the correlation with other
landmarks. The effect of change of projective coordinates due to
projective frame selection, can be understood via a group of
projective
transformations, but is beyond the scope of this paper.\\
We return to the reconstruction of a spatial configuration. Having
in view the definition \ref{prsh} of a projective shape of a
configuration, Theorem \ref{reconstruction} can be stated as
follows:
\begin{theorem}\label{key_result}
A spatial $\mathcal{R}$ reconstruction of a 3D configuration
$\mathcal{C}$ can be obtained in absence of occlusions from two of
its ideal camera views. Any such 3D reconstruction $\mathcal{R}$
of $\mathcal{C}$ has the same projective shape as $\mathcal{C}.$
\end{theorem}
\begin{remark}\label{essential-simplif}
Since the projective shape of the 3D reconstruction configuration
from a pair of images is uniquely determined, and since
multiplying by imposed internal camera parameters matrix keeps the
projective shape of the reconstruction unchanged, one may also fix
the internal camera parameters conveniently and estimate the
essential matrix instead of the fundamental matrix. An eight point
algorithm for estimation of the essential matrix is given in Ma
et. al. (2004, p. 121), for given internal parameters.
\end{remark}
\begin{remark} Another approach to projective shape has been recently
initiated by Kent and Mardia (2006, 2007). This approach has the
advantage of being invariant with respect to the group of
permutations of landmark indices, however it involves a nonlinear
approximation to the matrix solution of a data driven equation in
an $m \times m$ matrix, and has not been yet applied in projective
shape analysis for $m > 1.$
\end{remark}

\section{Nonparametric Estimation and Testing for the Projective Shape
a 3D Configuration} Assume $J: M \to \mathbb{R}^N$ is an embedding
of the $d$ dimensional complete manifold $M.$ Bhattacharya and
Patrangenaru (2003) defined the extrinsic mean $\mu_J$ of a
$J-$nonfocal random object ( r.o.) $Y$ on  $M$ by
\begin{equation}\label{extmean}
\mu_J =: J^{-1}(P_J(\mu)),
\end{equation}
where $\mu = E(J(Y))$ is the mean vector of $J(Y)$ and $P_J :
\mathcal{F}^c \to J(M)$ is the ortho-projection on $J(M)$ defined on
the complement of the set $\mathcal{F}$ of focal points of $J(M).$
 The extrinsic covariance matrix of $Y$ with respect to a
local frame field $y \rightarrow (f_1(y),\dots,f_d(y)),$  for
which $(dJ(f_1(y)),\dots ,dJ(f_d(y)))$ are orthonormal vectors in
$\mathbb{R}^N,$ was defined in Bhattacharya and Patrangenaru
(2005). If $\Sigma$ is the covariance matrix of $J(Y)$ (regarded
as a random vector on $\mathbb{R}^N),$ then $P_J$ is
differentiable at $\mu.$ In order to evaluate the differential
$d_\mu P_J$ one considers a special orthonormal frame field to
ease the computations. A local ortho-frame field $(e_1(p),e_2 (p)
,\dots, e_N(p))$ defined on an open neighborhood $U \subseteq
\mathbb{R}^N$ of $P_J(M)$ is {\it adapted to the embedding $J$ }
if $\forall y \in J^{-1}(U), (e_r(J(y))=d_yJ(f_r(y)), r = 1,
\dots, d$.
Let $e_1,e_2 , \dots, e_N$  be the canonical  basis of
$\mathbb{R}^N$ and assume $(e_1(p),e_2 (p) ,\dots, e_k(p))$ is an
adapted frame field around $P_J(\mu)=J(\mu_J).$ Then $\Sigma_E$
given by
\begin{eqnarray} \label{extco-man}
{\Sigma}_E = \left[ \sum^d_{a=1} d_\mu P_J (e_b) \cdot e_a(P_J
(\mu)) e_a (P_J (\mu)) \right]_{b=1,...,N} \Sigma \nonumber \\
\left[ \sum^d_{a=1} d_\mu P_J (e_b) \cdot e_a (P_J (\mu))
e_a(P_J(\mu)) \right]_{b=1,...,N}^T.
\end{eqnarray}
is the {\it extrinsic covariance matrix} of $Y$ with respect to
$(f_1(\mu_J),...,f_d(\mu_J)).$
The projective shape space $P\Sigma^k_m$ is homeomorphic to $M =
(\mathbb R P^m)^{q}, q = k - m -2 .$ $R P^m,$ as a particular \
case of a Grassmann manifold, is equivariantly embedded in the
space $S(m+1)$ of $(m+1) \times (m+1)$ symmetric matrices (
Dimitric (1996) ) via $ j: \mathbb R P^m \to S(m+1),$ given by
\begin{equation} \label{j}
j([x]) = x x^T.
\end{equation}
Patrangenaru (2001) and Mardia and Patrangenaru (2005) considered
the resulting equivariant embedding of the projective shape space
$$J = j_k: P\Sigma^k_m=(\mathbb{R}^m)^q \to (S(m+1))^q$$ defined by
\begin{equation} \label{j_k}
j_k([x_1],...,[x_q]) = (j[x_1],...,j[x_q]),
\end{equation}
where $x_s \in \mathbb R^{m+1},  x_s^T x_s = 1, \forall s =
1,...,q$.
\begin{remark}
The embedding $j_k$ in \eqref{j_k} yields the fastest known
computational algorithms in projective shape analysis. Basic axial
statistics related to Watson's method of moments such as the sample
mean axis ( Watson(1983)) and extrinsic sample covariance matrix
(Prentice(1984)) can be expressed in terms of $j_{m+3}=j$.
\end{remark}
A random projective shape $Y$ of a $k$-ad in $\mathbb{R}P^m$ is
given in axial representation by the multivariate random axes
\begin{equation} \label{multiaxes}(Y^1, \dots, Y^q), Y^s
= [X^s], (X^s)^TX^s =1, \forall s = 1, \dots, q = k -m -2.
\end{equation}
From Bhattacharya and Patrangenaru (2003) or Mardia and
Patrangenaru (2005) it follows that in this multivariate axial
representation of projective shapes, the extrinsic mean projective
shape of $(Y^1, \dots, Y^q)$ exists if $\forall s = 1, \dots, q,$
the largest eigenvalue of $E(X^s (X^s)^T)$ is simple. In this case
$\mu_{j_k}$ is given by
\begin{eqnarray} \label{meanprsh}
\mu_{j_k} = ([\gamma_1 (m + 1)],...,[\gamma_q (m + 1)])
\end{eqnarray}
where $\lambda_s(a)$ and $\gamma_s(a), a=1, \dots, m+1$ are the
eigenvalues in increasing order and the corresponding unit
eigenvector of $E(X^s (X^s)^T).$\\
If $Y_r, r = 1,\dots, n$ are i.i.d.r.o.'s from a population of
projective shapes ( in its multi-axial representation), for which
the mean shape $\mu_{j_k}$ exists, from a general consistency
theorem for extrinsic means on manifolds in Bhattacharya and
Patrangenaru (2003) it follows that the extrinsic sample mean
$[\overline{Y}]_{j_k,n}$ is a strongly consistent estimator of
$\mu_{j_k}.$ In the multivariate axial representation, $Y_r$ is
given by
\begin{eqnarray} \label{randprsh}
Y_r = ([X^1_{r}],\dots,[X^q_{r}]), (X^s_{r})^T X^s_{r} = 1; s
=1,...,q.
\end{eqnarray}
 Let $J_s$ be
the random symmetric matrix given by
\begin{eqnarray} \label{xxtbar}
J_s = n^{-1} \Sigma^n_{r=1} X^s_{r} (X^s_{r})^T ~~,~~ s = 1, \dots,
q,
\end{eqnarray}
and let $d_s(a)$ and $g_s(a)$ be the eigenvalues in increasing order
and the corresponding unit eigenvector of $J_s$, $a = 1, \dots,
m+1.$ Then the sample mean projective shape in its multi-axial
representation is given by
\begin{eqnarray} \label{samplmprsh}
\overline{Y}_{j_k,n} = ([g_1( m + 1)],\dots, [g_q (m + 1)]).
\end{eqnarray}
\begin{remark}
Some of the results in this section are given without a proof in
Mardia and Patrangenaru (2005). For reasons presented in Remark
\ref{corrections} we give full proofs of these results.
\end{remark}
To determine the extrinsic covariance matrix \eqref{extco-man} of
\eqref{multiaxes}, we note that the vectors
\begin{eqnarray} \label{tanprsh}
f_{(s,a)} = (0, \dots, 0, \gamma_s (a),0,\dots, 0),
\end{eqnarray}
with the only nonzero term in position $s, s\in \overline{1,q},\ a
\in \overline{1,m},$ yield a basis in the tangent space at the
extrinsic mean $T_{\mu_{j_k}}(\mathbb{R}P^m)^q,$ that is
orthonormal with respect to the scalar product induced by the
embedding $j_k.$ The vectors $e_{(s,a)}, \forall s\in
\overline{1,q}, \forall a \in \overline{1,m},$ defined as follows:
\begin{equation} \label{ortho} e_{(s,a)}=:d_{\mu_{j_k}}j_k(f_{(s,a)}).
\end{equation}
form an orthobasis of $T_{j_k(\mu_{j_k})}(\mathbb{R}P^m)^q.$ We
complete this orthobasis to an orthobasis of q-tuples of matrices
$(e_i)_{i \in \mathcal{I}}$ for $(S(m+1))^q,$ that is indexed by the
set $\mathcal{I},$ the first indices of which are the pairs $(s,a),
s = 1,\dots, q; a = 1, \dots, m$ in their lexicographic order.
Let $E_a^b$ be the $(m+1) \times (m+1)$ matrix with all entries
zero, except for an entry $1$ in the position $(a,b)$. The standard
basis of $S(m+1)$ is given by $e_a^b = E_a^b+E_b^a, 1 \leq a \leq b
\leq m+1$. For each $s = 1,...,q$ , the vector
$$
(_se_a^b) = (0_{m+1},...,0_{m+1},e_a^b,0_{m+1},...,0_{m+1})
$$
has all the components  zero matrices $0_{m+1} \in S(m+1)$, except
for the $s$-th component, which is the matrix $e_a^b$ of the
standard basis of $S(m+1,\mathbb{R})$; the vectors $_s e_a^b, s =1,
\dots, q, 1 \leq a \leq b \leq m+1$ listed in the lexicographic
order of their
indices $(s,a,b)$ give a basis of $S(m+1)^q$. \\
Let $\Sigma$ be the covariance matrix of $j_k(Y^1, \dots, Y^q)$
regarded as a random vector in $(S(m+1))^q,$ with respect to this
standard basis, and let $P =: P_{j_k}: (S(m+1))^q \to
{j_k}((\mathbb{R}P^m)^q)$ be the projection on
$j_k((\mathbb{R}P^m)^q).$ From \eqref{extco-man} it follows that
the
 extrinsic covariance matrix of $(Y^1, \dots, Y^q)$
with respect to the basis \eqref{tanprsh} of
$T_{\mu_{j_k}}(\mathbb{R}P^m)^q$ is given by
\begin{eqnarray} \label{extcov}
\Sigma_E = \left[ e_{(s,a)}(P(\mu))\cdot d_\mu P (_r e_a^b)
\right]_{(s=1,\dots, q),
(a=1,\dots, m)} \cdot \Sigma \nonumber\\
\cdot \left[ e_{(s,a)}(P(\mu))\cdot d_\mu P (_r e_a^b)
\right]^T_{(s=1,\dots, q), (a=1,\dots, m)}.
\end{eqnarray}
Assume $Y_1,\dots,Y_n$ are i.i.d.r.o.'s (independent identically
distributed random objects) from a $j_k$-nonfocal probability
measure on $(\mathbb{R}P^m)^q,$ and $\mu_{j_k}$ in
\eqref{meanprsh} is the extrinsic mean of $Y_1.$
We arrange the pairs of indices $(s,a), s = 1,\dots, q; a = 1,
\dots, m$, in their lexicographic order, and define the $(mq) \times
(mq)$ symmetric matrix $G_n$, with the entries
\begin{eqnarray} \label{extcovprsh}
{G_n}_{(s,a),(t,b)} = n^{-1}(d_s(m + 1) -  d_s(a) )^{-1} ( d_t(m+1)
-d_t(b) )^{-1}\cdot \nonumber \\
\cdot \sum^n_{r=1} ( g_s(a)^T  X^s_{r}) (g_t(b)^T X^t_{r}) (g_s(m +
1)^T X^s_{r}) (g_t(m + 1)^T  X^t_{r}) .
\end{eqnarray}
\begin{lemma}\label{lemma}
$G_n$ is the extrinsic sample covariance matrix estimator of
$\Sigma_E.$
\end{lemma}
\textbf{Proof}. The proof of lemma \ref{lemma} is based on the
equivariance of the embedding $j_k.$ As a preliminary step note that
the group $SO(m+1)$ acts as a group of isometries of $\mathbb R
P^m.$ If $R\in SO (m+1)$ and $[x] \in \mathbb R P^m$ then the action
$R([x]) = [Rx]$ is well defined. $SO(m + 1)$ acts by isometries also
on $S_+ (m+1,\mathbb R)$ via $R(A) = RAR^T$.  Note that the map
$j(x) = xx^T$ is equivariant since
$$
j (R[x]) = j ([Rx]) = (Rx) (Rx)^T = R j ([x]) R^T = R (j([x])).
$$
Therefore, for $q \geq 1$ the group $(SO (m + 1))^q$ acts as a
group of isometries of $(\mathbb R P^m)^q$ and also
\begin{equation} \label{actiononproj}
((R_1,...,R_q) \cdot (A_1,...,A_q) = (R_1A_1 R_1^T,...,R_q A_q
R_q^T), R_j \in SO (m+1), j = 1,\ldots,q.
\end{equation}
The map $j_k$ is equivariant with respect to this action since
\begin{eqnarray} \label{equiaction}
j_k((R_1,...,R_q)\cdot([x_1],...,[x_q]))
= (R_1,...,R_q) \cdot j_k([x_1],...,[x_q]), \nonumber \\
\forall (R_1,...,R_q) \in (SO(m+1))^q, \forall ([x_1],...,[x_q]) \in
({\mathbb{R}P^{m}})^q.
\end{eqnarray}
We set $M = j_k(({\mathbb{R}P^{m}})^q)$. Let $M_1^{m+1}$ be the set
of all matrices of rank 1 in $S_+(m+1)$. Note that $M$ is the direct
product of $q$ copies of $M_1^{m+1}$. Recall that
\begin{equation} \label{projectiononM}
P : ({S_+(m+1,\mathbb{R})})^q \rightarrow M
\end{equation}
is the projection on $M$. If $Y_r = ([X^1_r],\ldots,[X^q_r]),
r=1,...,n$, are i.i.d.r.o.'s from a probability distribution on
$({\mathbb{R}P^{m}})^q,$ we set
$$
V_r = j_k (Y_r).
$$
From the equivariance of $j_k$, w.l.o.g. ( without loss of
generality ) we may assume that $\overline{j_k(Y)}= \tilde{D} =
(\tilde{D}_1,\ldots,\tilde{D}_q)$ where $\tilde{D}_s \in
S_+(m+1,\mathbb{R})$ is a diagonal matrix, $s = 1,\ldots, q$.
Therefore
$$
\overline{Y}_{j_k,n} = ([g_1(m+1)],...,[g_q(m+1)])
$$
where $\forall s=1,...,q, \forall a=1,...,m + 1$, $g_a(s) = e_a$
are
the eigenvectors of  $\tilde{D}_s.$\\
It is obvious that if $\overline{V}$ is the sample mean of $V_r, r
= 1,\ldots, n,$ then
\begin{equation}
j_k (\overline{Y}_{j_k,n}) = P (\overline{V}) = P(\tilde{D}).
\end{equation}
Therefore w.l.o.g. we may assume that
\begin{equation} \label{gande}
g_s(a)=e_a, \forall s=1,\dots ,q, \forall a = 1, \dots, m+1,
\end{equation}
and that $j_k(p)=P(\overline{V})$ is given with $p =
([e_{m+1}],\ldots,[e_{m+1}])$.  The tangent space $T_p (\mathbb R
P^m)^q$ can be identified with $(\mathbb R^m)^q,$ and with this
identification $f_{(s,a)}$ in \eqref{tanprsh} is given by
$f_{(s,a)} = (0,\ldots, 0, e_a, 0,\ldots,0)$ which has all vector
components zero except for position $s,$ which is the vector $e_a$
of the standard basis of $\mathbb R^m$. We may then assume that
$e_{(s,a)} (\tilde{D}): = d_p j_k (_ie_s)$. From a straightforward
computation which can be found in Bhattacharya and Patrangenaru
(2005) it follows that ${d_{\tilde{D}}}P(_s e_a^b)=0$ , except for
\begin{equation} \label{dp} {d_{\tilde{D}}}P((_se_a^{m+1}))= \{ d_s(m+1)-d_s(a) \}
e_{(s,a)}(P_k(\tilde{D})).
\end{equation}
If $Y_r, r = 1, \dots, n$ is given by \eqref{randprsh}, from
\eqref{dp}, \eqref{extco-man}  we obtain
\begin{equation}
(G_n)_{(i,a),(j,b)}=n^{-1} \{ d_i(m+1)-d_i(a) \}^{-1}
\{d_j(m+1)-d_j(b)\}^{-1} \sum_{r}  {_{i}X_r^a} ~{}_jX_r^b ~~
{_{i}X_r^{m+1}}_jX_r^{m+1},
\end{equation}
which is \eqref{extcovprsh} expressed in the selected basis, thus
proving the Lemma $\blacksquare$\\
The proof of Theorem \ref{asymptprsh} is elementary following from
Lemma \ref{lemma}, and from the observation that $V_1$ has a
multivariate distribution with a finite covariance matrix $\Sigma$
since $(\mathbb{R}P^m)^q$ is compact. For $n$ large enough,
$\overline{V}$ has approximately a multivariate normal
distribution $\mathcal{N}(\mu, \frac{1}{n}\Sigma) $ and from the
delta method ( Fergusson 1996, p.45 ), it follows that
\begin{equation} \label{delta}
P( \overline{V}) \sim \mathcal{N}(P(\mu)=j_k(\mu_k), \frac{1}{n}
d_{\mu}P\Sigma d_{\mu}P^T).
\end{equation}
The range of the differential $d_\mu P$ is a subspace of
$T_{P(\mu)}j_k((\mathbb{R}P^m)^q),$ therefore the asymptotic
distribution of $P(\overline{V})$ is degenerate. If we decompose
$S(m+1))^q = T_{P(\mu)}j_k((\mathbb{R}P^m)^q)\oplus
T_{P(\mu)}j_k((\mathbb{R}P^m)^q)^\perp$ into tangent and normal
subspaces, then the covariance matrix of the tangential marginal
distribution of $tanP(\overline{V})$ is ${1\over n} \Sigma_E,$
which is nondegenerate because the generalized extrinsic
covariance is given by the determinant $det(\Sigma_E) =
\Pi_{s=1}^q\lambda _s(a),$ which is positive. Because
$\overline{V}$ is a strongly consistent estimator of $\mu,$ and
$S_n$ is a strongly consistent estimator of $\Sigma$, from
Slutsky's theorems (Fergusson, 1996, p.42) it follows that
 $ G_n $ in \eqref{extcovprsh} is a strongly consistent estimator of $\Sigma_E$.
Let $U=[(_sU_1,...,_sU_m)_{s=1,\dots,q}]^T$ be the random vector
whose components are the components of $tanP(\overline{V})$ w.r.t.
the basis $e_(s,a)(\tilde{D})$ which is given in the proof of
Lemma \ref{lemma}. Since $G_n$ is a consistent estimator of
$\Sigma_E,$ it follows that $Z_n = \sqrt{n} G_n^{-\frac{1}{2}}U$
converges to a ${\cal{N}}(0,{I_{mq}})$ - distributed random
vector, and $Z_n^T Z_n$ converges to a random variable with a
chi-square distribution with $mq$ degrees of freedom. If one uses
the equivariance again, one gets $Z_n^T Z_n =
T(\overline{Y}_{j_k,n};\mu)$ in \eqref{T}, which completes the
proof
of Theorem \ref{asymptprsh} $\blacksquare$\\
In preparation for an asymptotic distribution of
$\overline{Y}_{j_k,n}$ we set
\begin{equation} \label{d}
D_s = ( g_s(1),..., g_s(m) )\in {\cal M} (m+1,m; \mathbb R), s = 1,
\dots, q.
\end{equation}
If  $\mu = ([\gamma_1],\dots,[\gamma_q])$,  where $\gamma_s \in
\mathbb{R}^{m+1}, \gamma_s^T\gamma_s =1,$ for $ s = 1,\dots, q$, we
define a Hotelling's $T^2$ type-statistic
\begin{eqnarray} \label{T}
T(\overline{Y}_{j_k,n}; \mu) = n( \gamma^T_1 D_1,\dots,\gamma^T_q
D_q) G_n^{-1} (\gamma^T_1 D_1,\dots, \gamma^T_q D_q)^T.
\end{eqnarray}
\begin{theorem}\label{asymptprsh} Assume  $(Y_r)_{r = 1,\dots, n}$
are i.i.d.r.o.'s on $(\mathbb{R}P^m)^q$, and $Y_1$ is
$j_k$-nonfocal. Let $\lambda _s(a) \mbox{ and } \gamma_s(a)$ be
the eigenvalues in increasing order, respectively the
corresponding unit eigenvectors of $E[X_1^a (X_1^a)^T].$ If
$\lambda _s (1) > 0, \mbox{ for } s =1,\dots,q$, then
$T(\overline{Y}_{j_k,n}; \mu_{j_k})$ converges weakly to a
$\chi^2_{mq}$ distributed random variable.
\end{theorem}
If $Y_1$ is a $j_k$-nonfocal population on $(\mathbb R P^m)^q,$
since $(\mathbb R P^m)^q$ is compact, it follows that $j_k(Y_1)$
has finite moments of sufficiently high order. According to
Bhattacharya and Ghosh (1978), this, along with an assumption of a
nonzero absolutely continuous component, suffices to ensure an
Edgeworth expansion up to order $O(n^{-2})$ of the pivotal
statistic $T(\overline{Y}_{j_k,n}; \mu_{j_k}),$ and implicitly the
bootstrap approximation of this statistic.
\begin{corollary} \label{bootprsh}
Let $Y_r = ([X^1_r],\ldots,[X^q_r]), X_{st}^TX_{st}=1, s = 1,
\dots q, r = 1, \dots, n$, be i.i.d.r.o.'s from a $j_k$-nonfocal
distribution on $(\mathbb R P^m)^q$ which has a nonzero absolutely
continuous component, and with $\Sigma_E >0.$ For a random
resample with repetition $(Y^\ast_1,...,Y^\ast_n)$ from $(Y_1,
\dots, Y_n)$, consider the eigenvalues $d^{\ast}_s (a), a = 1,
\dots, m + 1$ of $\frac{1}{n} \sum^n_{r = 1} X^\ast_{rs} X^{\ast
T}_{rs}$ in their increasing order, and the corresponding unit
eigenvectors $g^{\ast}_s (a), a = 1, \dots, m + 1.$ Let $G_n^\ast$
be the matrix obtained from $G_n$, by substituting all the entries
with $\ast-$entries.  Then the bootstrap distribution function of
the statistic \begin{eqnarray} \label{bootT2}
T(\overline{Y}^\ast_{j_k}; \overline{Y}_{j_k}) = n(g_1(m + 1)
D^\ast_{1},\dots,
g_q(m + 1)D^\ast_q)~ G_n^{\ast -1} \nonumber \\
(g_1(m + 1) D^\ast_{1},\dots, g_q(m + 1)D^\ast_q)^T
\end{eqnarray}
approximates the true distribution of $T([\overline{Y}_{j_k};
\mu_{j_k}])$ given by \eqref{T}, with an error of order
$0_p(n^{-2})$.
\end{corollary}
\begin{remark}\label{corrections}
The above corollary is from Mardia and Patrangenaru (2005).
Formula \eqref{bootT2} in that paper has unnecessary asterisks for
$g_s(m+1),$ a typo that is corrected here. Also the condition
$\Sigma_E >0$ is missing there, as well as in their Theorem 5.1.
Another typo in Mardia and Patrangenaru (2005) is in their
definition of $\tilde{D}_s:$ the last column of $\tilde{D}_s$
 should not be there. The correct formula is
\eqref{d}. Note that $\tilde{D}_s = (D_s | g_s(m+1)).$
\end{remark}
Theorem \ref{asymptprsh} and Corollary \ref{bootprsh} are useful
in estimation and testing for mean projective shapes. We may
derive from \eqref{asymptprsh} the following large sample
confidence region for an extrinsic mean projective shape
\begin{corollary}\label{largeconfprsh} Assume  $(Y_r)_{r = 1,\dots, n}$
are i.i.d.r.o.'s from a $j_k-$nonfocal probability distribution on
$(\mathbb{R}P^m)^q,$ and $\Sigma_E >0.$ An asymptotic
$(1-\alpha)$-confidence region for $\mu_{j_k}=[\nu]$ is given by
$R_\alpha({\bf Y}) = \{[\nu]:T(\overline{Y}_{j_k,n};[\nu]) \leq
\chi^2_{mq,\alpha}\}$,
 where $T([\overline{Y}_{j_k},[\nu])$ is given in \eqref{T}. If the
 probability measure
 of $Y_1$ has
  a nonzero-absolutely continuous component w.r.t. the volume
  measure on $(\mathbb{R}P^m)^q,$ then the coverage error of
 $R_\alpha({\bf Y})$ is  of order $O_P(n^{-1}).$
\end{corollary}
\par For small samples the coverage error could be quite large, and
the bootstrap analogue in Corollary\ \ref{bootprsh} is preferable.
Consider for example the {\it one sample testing problem for mean
projective shapes}:
\begin{equation} \label{onesample}
H_0: \mu_{j_k} = \mu_0 \ \text{vs.} \ H_1: \mu_{j_k} \neq \mu_0.
\end{equation}
\begin{corollary} The large sample p-value for the testing problem
\eqref{onesample} is $p = Pr ( T > T(\overline{Y}_{j_k,n}; \mu_0)),$
where $T(\overline{Y}_{j_k,n}; \mu)$ is given by \eqref{T}.
\end{corollary}
\par In the small sample case, problem \eqref{onesample} can be answered based
on Corollary\ \ref{bootprsh} to obtain the following $100(1
-\alpha)\%$ bootstrap confidence region for $\mu_{j_k}:$
\begin{corollary} \label{onsamplesmall}Under the hypotheses of Corollary \ref{bootprsh},
The corresponding $100(1-\alpha)\%$ confidence region for
$\mu_{j_k}$ is
\begin{equation} \label{confbootalpha}
C^*_{n,\alpha}:=j_k^{-1}(U^*_{n,\alpha})
\end{equation}
with $U^*_{n,\alpha}$ given by
\begin{equation} \label{bootcutalpha}
U^*_{n,\alpha}=\{\mu \in j_k((\mathbb R P^m)^q):
T(\overline{y}_{j_k,n}; \mu) \leq c^*_{1-\alpha}\},
\end{equation}
where $c^*_{1-\alpha}$ is the upper $100(1-\alpha)\%$ point of the
values of $T(\overline{Y}^\ast_{j_k}; \overline{Y}_{j_k})$ given
by \eqref{bootT2}. The region given by
\eqref{confbootalpha}-\eqref{bootcutalpha} has coverage error
$O_P(n^{-2}).$
\end{corollary}
If $\Sigma_E$ is singular and all the marginal axial distributions
have positive definite extrinsic covariance matrices, one may use
simultaneous confidence ellipsoids to estimate $\mu_{j_k}.$ Assume
$(Y_r)_{r = 1,\dots, n}$  are i.i.d.r.o.'s from a $j_k-$nonfocal
probability distribution on $(\mathbb{R}P^m)^q.$ For each $s = 1,
\dots, q$ let $\Sigma_s$ be the extrinsic covariance matrix of
$Y^s_1,$ and let $\overline{Y}^s_{j,n}$ and $G_{s,n}$ be the
extrinsic sample mean and the extrinsic sample covariance matrix
of the $s$-th marginal axial. If the probability measure
 of $Y^s_1$ has a nonzero-absolutely continuous component w.r.t. the volume
  measure on $(\mathbb{R}P^m),$  and if for $s = 1, \dots, q$ and for $[\gamma_s]
\in \mathbb{R}P^m, \gamma_s^T\gamma_s =1,$ we consider the
statistics:
\begin{equation} \label{marginalT}
T_s= T_s(\overline{Y}^s_{j,n}, [\gamma_s]) = n \gamma_s^T D_s
G_{s,n}^{-1}D_s^T\gamma_s
\end{equation}
and the corresponding bootstrap distributions
\begin{equation} \label{bootTs}
T_s^\ast= T_s(\overline{Y}^{s\ast}_{j,n}, ; \overline{Y}^s_{j,n})
= n g_s(m + 1)^TD_s^\ast {G*_{s,n}}^{-1}D_s^{\ast T}g_s(m + 1).
\end{equation}
Since by Corollary \ref{asymptprsh} $T_s$ has asymptotically a
 $\chi^2_m$ distribution, we obtain the following
\begin{corollary} \label{onsamplesmall-s}
For $s = 1, \dots, q$ let $c^*_{s,1-\alpha}$ be the upper
$100(1-\alpha)\%$ point of the values of $T_s^\ast$ given by
\eqref{bootTs}. We set
\begin{equation} \label{confbootalpha-s}
C^*_{s,n,\beta}:=j_k^{-1}(U^*_{s,n,\beta})
\end{equation}
with $U^*_{s,n,\beta}$ given by
\begin{equation} \label{bootcutalpha-s}
U^*_{s,n,\beta}=\{\mu \in j(\mathbb R P^m):
T_s(\overline{y}^s_{j,n}; \mu) \leq c^*_{s,1-\beta}\}.
\end{equation}
If
\begin{equation}
R^{\ast}_{n,\alpha} = \cap_{s=1}^q C^*_{s,n,{\alpha\over q}},
\end{equation}
with  $C^*_{s,n,\beta},U^*_{s,n,\beta}$ given by
\eqref{confbootalpha-s}-\eqref{bootcutalpha-s}, then
$R^{\ast}_{n,\alpha}$ is a region of at least $100(1-\alpha)\%$
confidence for $\mu_{j_k}.$ The coverage error is of order
$O_P(n^{-2}).$
\end{corollary}
\begin{remark} \label{nonpivotal} If $\Sigma_E$ is singular, one may also use a method for constructing
nonpivotal bootstrap confidence regions for $\mu_{j_k}$ using
Corollary 5.1 of Bhattacharya and Patrangenaru (2003).
\end{remark}
\section*{ACKNOWLEDGEMENT}

The authors wish to thank the National Security Agency and the
National Science Foundation for their generous support. We would
also like to thank Rabi N. Bhattacharya and Adina Patrangenaru for
their suggestions that helped improve the manuscript.

\end{document}